\documentclass[12pt, a4paper]{article}
\usepackage{graphicx}
\usepackage{amssymb}
\usepackage{amsmath}
\usepackage{bm}
\usepackage{color}
\usepackage{cite}
\usepackage{epstopdf}            
\usepackage{epsfig}
            
\newcommand{\beq}{\begin{equation}}
\newcommand{\eeq}{\end{equation}}

\usepackage[colorlinks=true, linkcolor=black, citecolor=black,
urlcolor=black]{hyperref}

\begin{document}

\begin{titlepage}

\begin{flushright}

WU-HEP-16-03

\end{flushright}

\vskip 2cm
\begin{center}

{\Large
{\bf 
Diphoton excess at 750 GeV and LHC constraints \\
 in models with vector-like particles
}
}

\vskip 2cm

Junichiro Kawamura$^{1}$ and
Yuji Omura$^{2}$

\vskip 0.5cm

{\it $^1$Department of Physics, Waseda University, 
Tokyo 169-8555, Japan
}\\[3pt]
{\it $^2$
Kobayashi-Maskawa Institute for the Origin of Particles and the
Universe, \\ Nagoya University, Nagoya 464-8602, Japan}\\[3pt]

\vskip 1.5cm

\begin{abstract}
Recently, the ATLAS and CMS collaborations report excesses around $750$~GeV 
in the diphoton channels. This might be the evidence which reveals new physics beyond the Standard Model.
In this paper, we consider models with a $750$~GeV scalar and vector-like particles, which
couple each other through Yukawa couplings. 
The diphoton decay of the scalar is enhanced by the loop diagrams involving the extra colored particles. 
We investigate not only the setup required by the excesses, but also the LHC constraints, especially concerned with the vector-like particles. We consider the scenario that the extra colored particles decay to quarks and a dark matter (DM) via Yukawa couplings. Then, the signals from the vector-like particles are dijet, $b\overline{b}$ and/or $t\overline{t}$ with large missing energy. We discuss two possibilities for the setups: One is a model with vector-like fermions and a scalar DM, and the other is a model with vector-like scalars and a fermionic DM. 
We suggest the parameter region favored by the excess in the each case, and study the constraints 
 based on the latest LHC results at $\sqrt{s}=8$ TeV and $13$ TeV.
 We conclude that the favored region is almost excluded by the 
 LHC bounds, especially when the $750$ GeV scalar dominantly decays to DMs.
 The mass differences between the vector-like particles and the DM should be less than ${\cal O}(100)$ GeV (${\cal O}(10)$ GeV) to 
 realize the large diphoton signal and the large decay width, if the extra colored particle only decays to a top (bottom) quark and a DM. Otherwise, these scenarios are already excluded by the latest LHC results.  
\end{abstract}

\end{center}
\end{titlepage}

\section{Introduction}
\label{sec:intro}

Recently, the ATLAS and CMS collaborations have announced the results in the diphoton resonance 
searches at $\sqrt{s}=13$ TeV~\cite{diphotonexcess-ATLAS, diphotonexcess-CMS}. 
The ATLAS collaboration reports an excess around 750 GeV diphoton invariant mass region, based on 3.2 fb$^{-1}$ of $pp$ collisions. Assuming about $45$ GeV decay width, the local (global) significance is 3.9 (2.3) $\sigma$ ~\cite{diphotonexcess-ATLAS}. In the analysis under the narrow width hypothesis, the significance goes down to 3.6 (2.0) $\sigma$ ~\cite{diphotonexcess-ATLAS}.
Based on 2.6 fb$^{-1}$ data, the CMS collaboration also finds a rather mild excess of $\gamma \gamma$ events, peaked at 760 GeV. Assuming the narrow width, the local (global) significance is 2.6 (1.2) $\sigma$ ~\cite{diphotonexcess-CMS}. 
They are not still conclusive, but many possibilities of the 
diphoton resonances have been already proposed after the reports~\cite{diphoton-composite,diphoton-dm,Franceschini:2015kwy,diphoton-new,diphoton-axion,diphoton-stability,diphoton-susy,diphoton-vectorlike,diphoton-u1,diphoton-gut,
Dorsner:2016ypw,diphoton-ed,diphoton-2HDM,diphoton-higgcision,diphoton-gut2,diphoton-photoncollision}. 
The signal requires a large cross section, $\sigma (pp \to \gamma \gamma ) \simeq 10$ [fb],
as well as the $750$~GeV particle, so that it seems to suggest some extra particles around the scale,
which can enhance the cross section. 

One interesting possibility is to consider an extended Standard Model (SM) with vector-like particles, whose masses are less than ${\cal O}(1)$ TeV. In addition, we introduce a scalar ($S$) with $750$ GeV mass and $S$ has Yukawa couplings with the extra vector-like particles. 
The extra vector-like particles are colored and carry electro-magnetic charges. 
Then, $S$ can decay to $\gamma \gamma$ and $gg$ via the loop diagrams
involving the vector-like particles. 
This kind of scenarios have been already proposed, motivated by the excesses, and successfully reproduces the 
signal \cite{diphoton-vectorlike,diphoton-u1,diphoton-gut,Dorsner:2016ypw}, although large Yukawa couplings and 
very light vector-like particles are necessary.
Especially, the vector-like masses should be less than 1 TeV, and maybe confront the severe bounds from the searches for extra colored particles at the LHC. 
Then, we have to seriously check the consistency of the signal with the latest bounds on the exotic colored particles. These models have been discussed, motivated by the dark matter (DM) physics and baryon asymmetry 
\cite{vectorlike,vectorlike2}, 
so that it is worthwhile to investigate the current status of the models as well.

In this paper, we consider two types of scenarios: One is that the vector-like pairs are fermions and the other is
that the vector-like particles are bosons, namely squarks. 
In both cases, we also introduce a DM, which couples with
the SM quarks and the extra quarks or squarks through Yukawa couplings.
In order to write the Yukawa couplings, the DM should be bosonic in the extra-quark case and fermionic in the squark case. Then, the extra quarks/squarks, enhancing the diphoton signal, decay to the DM and the SM quarks. 
This setup predicts the signal with $jj/b\overline{b}/t\overline{t}$ plus large missing energy at the LHC. Besides,
the mono-jet signal may play a crucial role in test of our model. This is because 
the excesses may suggest ${\cal O}(10)$ GeV decay width of $S$ according to the ATLAS report, so that $S$ might decay to two DMs mainly.
Our main motivation of this paper is to study the collider bounds on the vector-like particles in the two scenarios, and investigate the constraints in the parameter region favored by the diphoton excesses. 

In Sec. \ref{sec;setup}, we introduce our setups in both cases and discuss
how to achieve the diphoton excesses in Sec. \ref{sec;signal}.  
The experimental bounds especially from the recent LHC results are studied in Sec. \ref{sec;bound},
and the allowed parameter region is summarized. Then we discuss the consistency of the favored region by the excess with the constraints from the new physics search in the $jj/b\overline{b}/t\overline{t}$ plus large missing energy channels.
Sec. \ref{sec;conclusion} is devoted to the conclusion.

\section{Setup}
\label{sec;setup}
\subsection{Type-I scenario}
\label{sec;type1}
In this section, we introduce our models. 
First, let us discuss the setup in the type-I scenario that the extra vector-like particles are fermonic and 
the DM is bosonic. 

We introduce $N_u$ extra up-type quarks, $U_\alpha$ ($\alpha=1,\dots, N_u$), and $N_d$ extra down-type quarks, $D_a$  ($a=1,\dots, N_d$).\footnote{Note that we can consider additional quarks charged under $\text{SU}(2)_L$, but they are not so effective to enhance diphoton signals because of the suppression from $1/6$ $\text{U}(1)_Y$ charges.  }
They are Dirac fermions and carry SM charges as in Table \ref{table1}.
\begin{table}[t]
\caption{Extra vector-like quarks in the type-I scenario. $\alpha$ and $a$ denote the flavors: $\alpha=1,\dots, N_u$ and $a=1,\dots, N_d$. }
 \label{table1}
\begin{center}
\begin{tabular}{ccccc}
\hline
\hline
Fields    & ~~$\text{SU}(3)_c$~~ & ~~$\text{SU}(2)_L$~~  & ~~$\text{U}(1)_Y$~~ &~~$Z_2$~~     \\ \hline  
  $U_{\alpha}$  & ${\bf 3}$         &${\bf1}$      &       $2/3$   & $-$         \\  
   $D_{ a}$  &  ${\bf 3}$        &${\bf1}$      &         $-1/3$      & $-$          \\  \hline 
   $\widetilde X$  & ${\bf1}$         &${\bf 1}$      &            $0$   & $-$     \\   
      $S$  & ${\bf1}$         &${\bf 1}$      &            $0$   & $+$     \\ \hline \hline
\end{tabular}
\end{center}
\end{table} 
Note that we assign $Z_2$ odd charges to $U_\alpha$ and $D_a$. In addition, we introduce a complex
scalar DM candidate $\widetilde X$ and a scalar $S$ with $750$ GeV mass.
The stability of $\widetilde X$ is guaranteed by the additional $Z_2$ symmetry.

According to the charge assignment in Table \ref{table1}, we can write down the potential for the extra quarks and scalars,
\begin{eqnarray}
V_{{\rm I}}&=&V_m+V_{X} + V_{\rm S}(S, \widetilde X,H), \\
&&V_m= (m_U + y_U S )_{\alpha \beta} \overline{U}^{\alpha}U^\beta +(m_D + y_D S )_{ab} \overline{D}^aD^b,   \\
&&V_X=(\lambda_u)_{\alpha \, i} \overline{U_L}^{\alpha} \widetilde X u^i_{R}+  (\lambda_d)_{a \, i} \overline{D_L}^{a} \widetilde X d^i_{R}+ h.c.. \label{eq;VX}
\end{eqnarray}
$u^i_{R}$ ($d^i_{R}$) are the right-handed up-type (down-type) quarks: $(u^1_{R}, \, u^2_{R}, \, u^3_{R})=(u_R, \, c_R, \, t_R)$ and $(d^1_{R}, \, d^2_{R}, \, d^3_{R})=(d_R, \, s_R, \, b_R)$.
The each coupling depends on $\alpha$, $a$, and/or $i$.
For simplicity, the alignment is assumed in Sec. \ref{sec;bound}.
Note that $V_{\rm S}$ is the potential for the scalars including the SM Higgs ($H$).
It is not explicitly written down here, but assumed that it includes the following 3-point couplings as well,
\beq
V_S \supset A_{XS} S \widetilde X^\dagger \widetilde X +h.c..
\eeq

\subsection{Type-II model}
\label{sec;type2}
Next, we consider the other possibility that the extra vector-like pairs are complex scalars charged under SU(3)$_c$ and U(1)$_Y$. They are denoted by $\widetilde U_\alpha$ and $\widetilde D_a$ and called squarks.
We also introduce a fermionic DM candidate, $X$, which is a Dirac fermion.
The charge assignment is summarized in Table \ref{table2}.
\begin{table}[t]
\caption{Extra vector-like squarks in the type-II scenario. $\alpha$ and $a$ denote the flavors: $\alpha=1,\dots, \widetilde N_u$ and $a=1,\dots, \widetilde  N_d$. }
 \label{table2}
\begin{center}
\begin{tabular}{ccccc}
\hline
\hline
Fields    & ~~$\text{SU}(3)_c$~~ & ~~$\text{SU}(2)_L$~~  & ~~$\text{U}(1)_Y$~~ &~~$Z_2$~~     \\ \hline  
  $\widetilde U_{\alpha}$  & ${\bf 3}$         &${\bf1}$      &       $2/3$   & $-$         \\  
   $\widetilde D_{ a}$  &  ${\bf 3}$        &${\bf1}$      &         $-1/3$      & $-$          \\  \hline 
   $ X$  & ${\bf1}$         &${\bf 1}$      &            $0$   & $-$     \\   
      $S$  & ${\bf1}$         &${\bf 1}$      &            $0$   & $+$     \\ \hline \hline
\end{tabular}
\end{center}
\end{table} 
Based on the charge assignment in Table \ref{table2}, we can write down the couplings for the extra matters:
\begin{eqnarray}
V_{{\rm II}}&=&\widetilde V_m+\widetilde V_{X} + \widetilde V_{\rm S}(S, X,H), \\
&& \widetilde V_m= (\widetilde m^2_U + A_U  S )_{\alpha \beta} \widetilde U^\dagger_{\alpha} \widetilde U_\beta +(\widetilde m^2_D + A_D S )_{ab} \widetilde D^\dagger_a \widetilde D_b,   \\
&&\widetilde V_X=( \widetilde \lambda_u)_{\alpha \, i} \widetilde U^{\alpha}_L \overline{X_L} u^i_{R}+  (\widetilde \lambda_d)_{a \, i} \widetilde D^{a \, \dagger } \overline{ X_L} d^i_{R}+ h.c..
\end{eqnarray}
Note that $(A_U)_{\alpha \beta}$ and $(A_D)_{ab}$ are the dimensional parameters.

\section{The diphoton signals}
\label{sec;signal}
In the both scenarios, $S$ does not directly couple with the SM quarks,
so that $S$ can be mainly produced by the gluon fusion.
Then $S$ decays to not only $gg$ but also $\gamma \gamma$ via the loop diagrams involving
the extra colored particles.
After the announcement of the diphoton excesses at the LHC, 
the signal for $gg \to S \to \gamma \gamma$ has been well studied.
Following Ref.~\cite{Franceschini:2015kwy}, 
the production of $S$, where the gluon fusion is dominant,
can be described in terms of the decay widths of $S  \to gg$
and $S \to \gamma\gamma$ and 
the integral of parton distribution functions of gluons ($C_{gg}$) 
by
\begin{equation}
\sigma(gg\to S \to \gamma\gamma)=
\frac{C_{gg}}{s \, m_{S}\, \Gamma_\textrm{tot}} \Gamma (S \to gg) \,
\Gamma(S \to \gamma\gamma).
\end{equation}
$C_{gg}$ is numerically given in Ref.~\cite{Franceschini:2015kwy}: 
$C_{gg}=2137$ at $\sqrt s =$13 TeV and $C_{gg}=174$ at $\sqrt s =$8 TeV.
$m_S$ and $\Gamma_{\textrm{tot}}$ are the mass and the total decay width of $S$.

The decay rates of $S$ to two gluons and two photons 
depend on the scenarios. In the type-I model, the partial decay widths 
are given as follows:
\begin{eqnarray}
\Gamma (S \to gg) &=&
\frac{\alpha_s^2 m^3_{S}}{128 \pi^3 }
\left| \sum^{N_u}_{\alpha=1} \frac{ y^{\alpha\alpha}_U}{m_{U \, \alpha}} {\cal A}_{1/2}^H (\tau_{\alpha})+ \sum^{N_d}_{a=1}  \frac{ y^{aa}_D }{m_{D \, a}} {\cal A}_{1/2}^H (\tau_{a})
\right|^2,
\\
\Gamma (S \to \gamma\gamma ) &=&
\frac{\alpha^2 m^3_{S} }{256 \pi^3 }
\left|  \sum^{N_u}_{\alpha=1}  \frac{y^{\alpha \alpha}_U}{m_{U \, \alpha}} N_c  \left (\frac{2}{3} \right )^2 {\cal A}_{1/2}^H (\tau_{\alpha})+  \sum^{N_d}_{a=1}  \frac{y^{a a}_D}{m_{D \, a}} N_c  \left (-\frac{1}{3} \right )^2 {\cal A}_{1/2}^H (\tau_{a})
\right|^2, 
\nonumber \\
&&
\end{eqnarray}
where $\alpha$ and $\alpha_s$ are the fine structure constant and the strong coupling. $\tau_\alpha $ and $\tau_a$ are given by $\tau_\alpha = m_{S}^2/4 (m^\alpha_U)^2$ and $\tau_a = m_{S}^2/4 (m^a_D)^2$, respectively.
Here we assume $m_{ U}$ and $m_{D}$ are diagonal: $(m_{U})_{\alpha \beta}=m^\alpha_U \delta_{\alpha \beta}$ and $(m_{D})_{a b}=m^a_D \delta_{a b}$. The loop function ${\cal A}_{1/2}^H (\tau)$ is defined by
\begin{eqnarray}
{\cal A}_{1/2}^H (\tau) = 2 [ \tau + (\tau - 1) f(\tau) ] / \tau^2,
\end{eqnarray}
where the fucntion $f(x)$ is given by
\begin{equation}
f(x) = \left\{ 
\begin{array}{ccc}
\arcsin^2 \sqrt{x} &,& \textrm{for } x \le 1;
\\
\displaystyle
-\frac{1}{4}\left[\log \frac{1+\sqrt{1-1/x}}{1-\sqrt{1-1/x}}
-i\pi \right]^2
&,& \textrm{for } x > 1.
\end{array}
\right.
\end{equation}

In the type-II case, the decay rates to $gg$ and $\gamma \gamma$ are induced by the scalar loops:
\begin{eqnarray}
\Gamma (S \to gg) &=&
\frac{\alpha_s^2 m^3_{S}}{128 \pi^3 }
\left| \sum^{N_u}_{\alpha=1} \frac{ A^{\alpha \alpha}_U }{ 2\widetilde m^2_{U \, \alpha} }  {\cal A}_{0}^H (\tau_{\alpha})+ \sum^{N_d}_{a=1} \frac{A^{aa}_D}{2 \widetilde m^2_{D \, a} } {\cal A}_{0}^H (\tau_{a})
\right|^2,
\\
\Gamma (S \to \gamma\gamma ) &=&
\frac{\alpha^2 m^3_{S} }{256 \pi^3 }
\left|  \sum^{N_u}_{\alpha=1}  \frac{A^{\alpha \alpha}_U }{ 2 \widetilde m^2_{U \, \alpha} } N_c  \left (\frac{2}{3} \right )^2 {\cal A}_{0}^H (\tau_{\alpha})+  \sum^{N_d}_{a=1}  \frac{ A^{aa}_D }{ 2 \widetilde m^2_{D \, a} } N_c  \left (-\frac{1}{3} \right )^2 {\cal A}_{0}^H (\tau_{a})
\right|^2.
\nonumber \\
&&
\end{eqnarray}
where ${\cal A}_0^H$ is given by
\begin{eqnarray}
{\cal A}_0^H(\tau) &=& -[\tau - f(\tau)]/\tau^2.
\end{eqnarray}

In principle, $S$ decay to the extra particles, as well as 
the dark matter particles.
Now, let us simply define the extra decay width $\Delta \Gamma$ (GeV) and the total decay 
width ($\Gamma_{\rm tot}$) of $S$  could be given by
\beq
\Gamma_{\rm tot}=\Delta \Gamma+\Gamma (S \to gg)+\Gamma(S \to \gamma \gamma).   
\eeq
It seems that the diphoton excess requires ${\cal O}(10)$-GeV $\Gamma_{\rm tot}$ and  ${\cal O}(10)$-fb diphoton signal at $\sqrt s =13$ TeV. It may be very difficult to consider only $S \to gg$ and $\gamma \gamma$,
in order to reproduce the excess. In these minimal setups, nonzero $\Delta \Gamma$ can be realized by 
the invisible decay: $S \to \widetilde X \widetilde X^\dagger$ or $X \overline X$.
We also give a study on the bound from the invisible decay.

\subsection{Constraints from the resonance searches at $\sqrt{s}=8$ TeV}
\label{subsec;LHC8}
Before showing our results on the diphoton excesses, let us summarize the constraints
relevant to the signal. The LHC experiments have surveyed the heavy resonances in the diboson channels,
at $\sqrt{s}=8$ TeV. 
\begin{table}[t]
\caption{The upper bounds on the diboson signals of the $750$ GeV resonance from the LHC results at $\sqrt{s}=8$ TeV. In the diphoton search, the CMS and ATLAS results are shown. The CMS result assumes the narrow width ($\Gamma_{\rm tot}=0.1$ GeV).   }
 \label{table3}
\begin{center}
\begin{tabular}{ccc}
\hline
\hline
channels    & ~~  upper bounds at  $\sqrt{s}=8$ TeV~~  \\ \hline  
  $\sigma (pp \to S \to gg)$        & 2 pb  \cite{CMS-dijet}   \\ 
  $\sigma (pp \to S \to \gamma \gamma)$      &  1.3 fb \cite{CMS-diphoton8} 2.2 fb \cite{Aad:2015mna}    \\  
   $\sigma (pp \to S \to Z \gamma)$  &  3.8 fb \cite{Aad:2014fha}       \\ 
   $\sigma ( pp \to S \to ZZ)$  &    22 fb  \cite{Aad:2015kna}       \\
      $\sigma ( pp \to S \to WW)$  &    40 fb  \cite{Aad:2015agg}      \\
      $\sigma ( pp \to S \to hh)$  &    10 fb \cite{Aad:2015uka}        \\  \hline \hline
\end{tabular}
\end{center}
\end{table} 
$S$ can decay to $ZZ$ and $Z \gamma$, as well as $\gamma \gamma$ and $gg$,
via the loop diagrams. $\Gamma (S \to ZZ)$ is suppressed by $(\sin \theta_W/\cos \theta_W)^4$,
compared to $\Gamma (S \to \gamma \gamma)$. $\Gamma (S \to Z \gamma)$ is also relatively small, so that the stringent constraints are from the diphoton and dijet searches at $\sqrt{s}=8$ TeV.
The relevant constraints are summarized in Table \ref{table3}.
In the next section, we draw the lower bounds on the vector-like particle masses. They would be around $300$-$600$ GeV, depending on the Yukawa couplings, as we see in Table \ref{table4} and Table \ref{table5}.

\subsection{Results on the diphoton signal}
\label{subsec;signal}
\begin{figure}[!t]
{\epsfig{figure=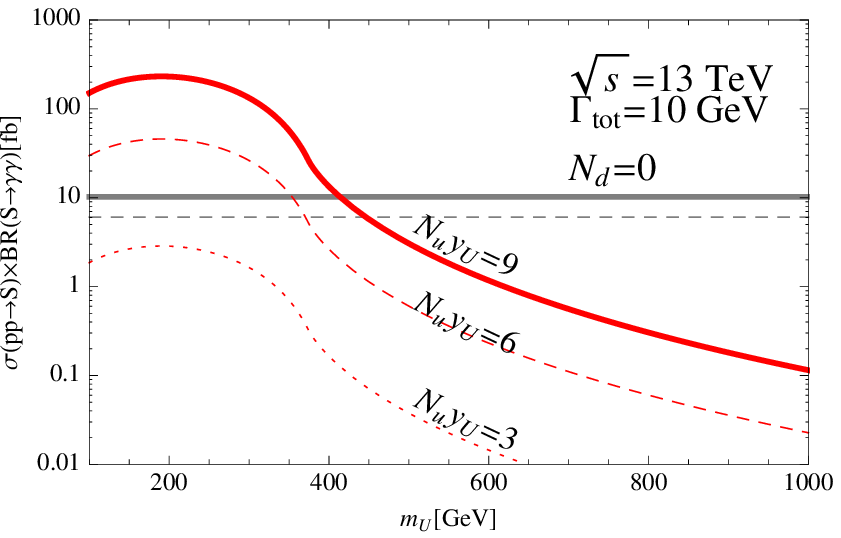,width=0.5\textwidth}}
{\epsfig{figure=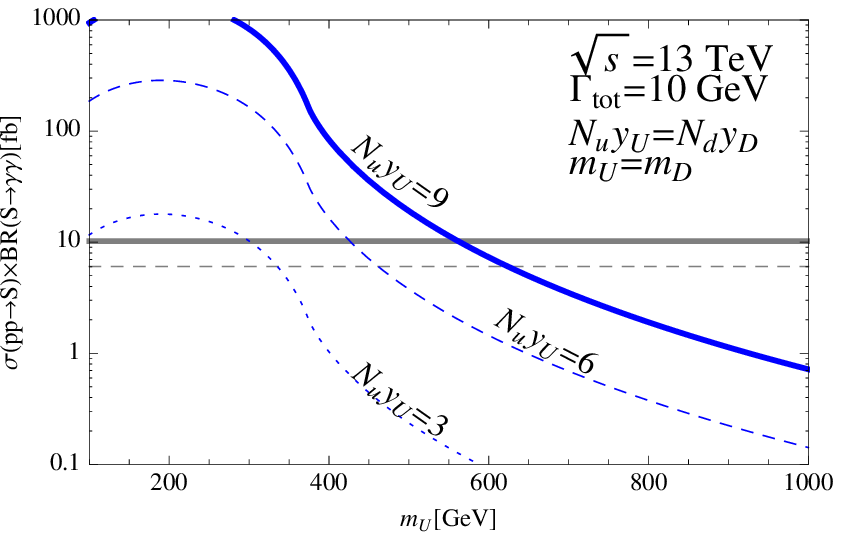,width=0.5\textwidth}}
\vspace{-0.5cm}
\caption{
$m_U$ vs. the diphoton signal via the gluon-gluon fusion at $\sqrt s =$13 TeV, in the type-I model with
$N_d=0$ (left) and $N_u y_U=N_d y_D$ (right). 
The total decay width is $10$ GeV. 
The each line corresponds to the prediction of the 750 GeV $S$, at $N_u y_U=3, \, 6, \, 9$, respectively. 
The thick (dashed) gray line is the exclusion line from the diphoton search at the ATLAS (CMS) experiment\cite{CMS-diphoton8, Aad:2015mna}.  }
\label{fig1}
\end{figure}

Finally, let us discuss the parameter region favored by the diphoton excess at the LHC in the Type-I model.
Simply, we assume that the extra quark masses are degenerate and the Yukawa couplings also do not depend on the flavors: $y^{\alpha \beta}_U=y_U \delta^{\alpha \beta}$ and $y^{a b}_D=y_D \delta^{ab}$. Fig. \ref{fig1} shows $\sigma (pp \to S) \times$BR$(S \to \gamma \gamma)$ via the gluon fusion on the each extra quark mass.
$m_S$ and $\Gamma_{\rm tot}$ are set to $m_S=750$ GeV and  $\Gamma_{\rm tot}=10$ GeV.
The each line corresponds to the case with $N_u y_U=3$ (dotted), $6$ (dashed), and $9$ (thick).
In the left panel, $N_d$ is vanishing and in the right, $N_d y_D=N_u y_U$ is assumed together with $m_D=m_U$.
The thick (dashed) gray line is the exclusion line from the diphoton search at the ATLAS (CMS) experiment\cite{CMS-diphoton8, Aad:2015mna}. 

The diphoton excess requires ${\cal O}(10)$-fb cross section at $\sqrt{s}=13$ TeV. As we see Fig. \ref{fig1}, such a large cross section can be realized, if $m_U$ is $300$-$600$ GeV and $y_U$ is ${\cal O} (1)$ when $N_u$ is $3$. 
The lower bounds on $m_U$ from the diphoton resonance search at $\sqrt s=8$ TeV are shown 
in Table \ref{table4}. The required cross section for the excess is close to the upper bound from the diphoton signal at $\sqrt s=8$ TeV, so that the lower values of $m_U$ are close to the required ones for the diphoton excess at $\sqrt s=13$ TeV.
Note that the lower bounds from dijet and $Z \gamma$ resonance searches are described in the brackets in Table \ref{table4}, although they are weaker. Such a light extra quark might be already excluded by the LHC experiments, so that
we have to investigate the consistency with the extra quark search.
\\

Similarly, we discuss the diphoton signals in the type-II scenario.
The enhancement via the squarks is, however, much less, so that it is very difficult to achieve the ${\cal O}(10)$-fb signal.
Let us simply assume $A^{\alpha \beta}_U=A_U \delta^{\alpha \beta}$ and $A^{a b}_D=A_D \delta^{ab}$.
Now we define $\widetilde y_U$ and $\widetilde y_D$ as $A_U= \widetilde y_U \widetilde m_U$ and $A_D= \widetilde y_D \widetilde m_D$, assuming $(\widetilde m^2_U)^{\alpha \beta}= \widetilde m^2_U \delta^{\alpha \beta}$ and $(\widetilde m^2_D)^{a b}=\widetilde m^2_D \delta^{ab}$.
Fig. \ref{fig2} shows $\sigma (pp \to S) \times$BR$(S \to \gamma \gamma)$ via the gluon fusion on the each extra squark mass, in the cases with vanishing $N_d$ (left) and $\widetilde N_d \widetilde y_D= \widetilde N_u \widetilde y_U$ (right). $m_S$ and $\Gamma_{\rm tot}$ are set to $m_S=750$ GeV and  $\Gamma_{\rm tot}=10$ GeV.
The each line corresponds to the case with $\widetilde N_u \widetilde y_U=10$ (dotted), $20$ (dashed), and $30$ (thick). As we see Fig. \ref{fig2}, very large $\widetilde N_u \widetilde y_U$ may be able to reach the ${\cal O} (10)$-fb cross section. When $\widetilde N_u$ and $\widetilde N_d$ are set to ${\cal O}(1)$, $\widetilde y_U$
and $\widetilde y_D$ might be safe for the perturbativity because they are still less than $4 \pi$,
but the Landau poles will appear at very low scales. 
Besides, the favored region also requires $\widetilde m_U$ to be from $200$ GeV to $400$ GeV,
which may be strongly constrained by the LHC experiments. 
The lower bounds on $\widetilde m_U$ from the diphoton channel at $\sqrt s=8$ TeV are shown 
in Table \ref{table5}. The bounds from dijet resonance search are described in the bracket, although they are weaker.

Based on the discussion here, we study the constraints from the surveys of extra colored particles in
the $jj$/$b\overline{b}$/$t\overline{t}$ signals with large missing energy at the LHC in the next section.

\begin{figure}[!t]
{\epsfig{figure=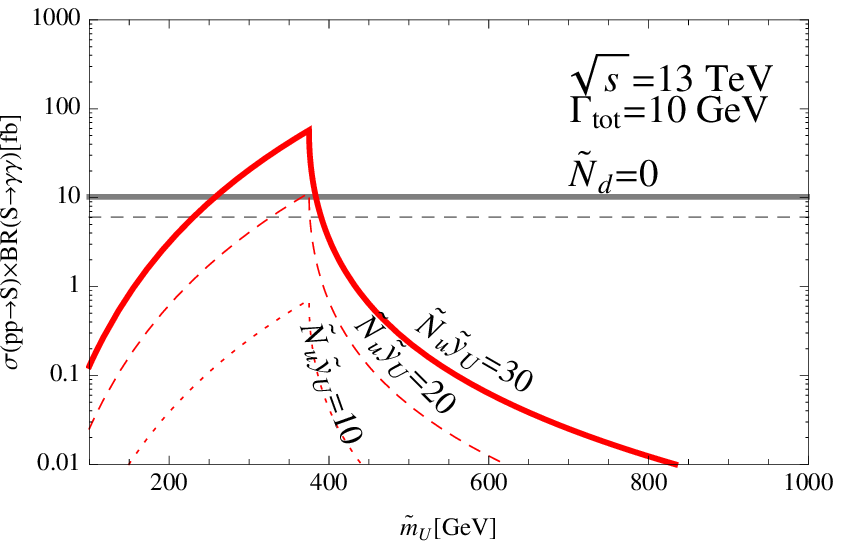,width=0.5\textwidth}}
{\epsfig{figure=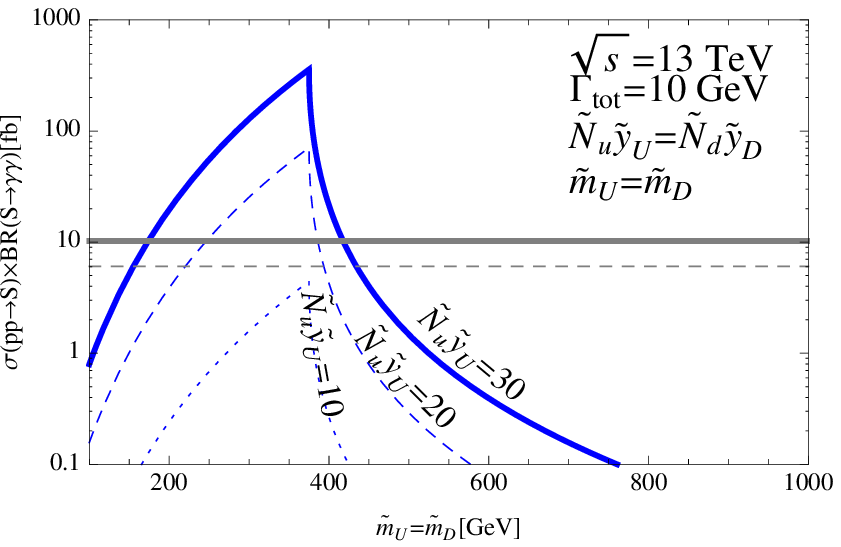,width=0.5\textwidth}}
\vspace{-0.5cm}
\caption{
$\widetilde m_U$ vs. the diphoton signal via the gluon-gluon fusion at $\sqrt s =$13 TeV, in the type-II model with
$\widetilde N_d=0$ (left) and $\widetilde N_u \widetilde y_U=\widetilde N_d \widetilde y_D$ (right). 
$\widetilde y_U$ ($\widetilde y_D$) is defined as $\widetilde y_U=A_U/\widetilde m_U$ ($\widetilde y_D=A_D/\widetilde m_D$).
The total decay width is $10$ GeV. 
The each line corresponds to the prediction of the 750 GeV $S$, at $\widetilde N_u \widetilde y_U=10, \, 20, \, 30$, respectively. 
The thick (dashed) gray line is the exclusion line from the diphoton search at the ATLAS (CMS) experiment \cite{CMS-diphoton8, Aad:2015mna}.  }
\label{fig2}
\end{figure}

\begin{table}[t]
\caption{The lower bounds on $m_U$ from the diphoton search at $\sqrt{s}=8$ TeV \cite{Aad:2015mna} in the type-I model.
The bounds from the dijet and $Z\gamma$ channels are also described in the brackets.   }
 \label{table4}
\begin{center}
\begin{tabular}{ccc}
\hline
\hline
 models & $N_u y_U$   & ~~  lower bounds on $m_U$~~  \\ \hline  
  $N_d=0$     & 9   &   414 (376) GeV  \\ 
      &  6  & 354 (295) GeV  \\   \hline
 $N_u=N_d$   &  9      &  562 (540) GeV  \\
     &  6  &  426 (414) GeV    \\
   &  3  & 300 (279) GeV   \\  \hline \hline
\end{tabular}
\end{center}
\end{table} 


\begin{table}[t]
\caption{The lower bounds on $\widetilde m_U$ from the diphoton search at $\sqrt{s}=8$ TeV \cite{Aad:2015mna} in the type-II model.
The bounds from the dijet channel are also described in the brackets. 
Note that the $S \to Z \gamma$ bound is weaker. }
 \label{table5}
\begin{center}
\begin{tabular}{ccc}
\hline
\hline
 models & $\widetilde N_u \widetilde y_U$   & ~~  lower bounds on $\widetilde m_U$~~  \\ \hline  
  $\widetilde N_d=0$     & 30   &   384 (376) GeV  \\ 
      &  20  & 375 GeV  \\   \hline
 $\widetilde N_u=\widetilde N_d$   &  30     &  418 (412) GeV  \\
     &  20  &  386 (383) GeV    \\  \hline \hline
\end{tabular}
\end{center}
\end{table} 


\section{Study on the LHC bounds on vector-like particles}
\label{sec;bound}
\subsection{Type-I model}
\label{subsec;type1}
First, let us investigate the constraint on the scalar DM at the LHC.
In our model, the 750 GeV scalar decays to the DMs as well as two photons and two gluons.
The observed excess, especially reported by the ATLAS collaboration, suggests ${\cal O}(10)$-GeV decay width
of $S$, so that $S$ mainly decay to the two DMs, once it is produced by the gluon fusion.
Then, the signal can be found, associated with a energetic jet. 
In fact, the ATLAS and CMS collaborations survey the mono-jet signals and 
draw the constraints on the DM mass and interaction \cite{Aad:2015zva,Khachatryan:2014rra}.

To estimate bounds from the mono-jet search, we adopt the following effective interaction: 
\begin{eqnarray}
\mathcal{L}_{ggS} = - \sum_{f=U_\alpha ,D_a} \frac{\alpha_s y_f N_f}{16\pi m_f} \mathcal{A}_{1/2}^H (\tau_f) S G^{a\mu\nu} G^a_{\mu\nu}, 
\end{eqnarray}
then a coefficient of the effective operator $ \widetilde X^\dagger \widetilde X \alpha_s (G_{\mu\nu}^a)^2$ can be written as, 
\begin{eqnarray}
\label{eq:Mstar}
\frac{1}{M^2_*} \equiv \frac{A_{XS}}{m_S^2} \sum_{f=U_\alpha ,D_a} \frac{\alpha_s y_f N_f}{16\pi m_f} \mathcal{A}_{1/2}^H (\tau_f) . 
\end{eqnarray}

If the DM mass $m_{\widetilde X}$ is smaller than $m_S/ 2$, the decay $S \rightarrow \widetilde X^\dagger \widetilde X$ is allowed, then it can contribute to the total decay width.
The tree-level partial decay width can be written as 
\begin{eqnarray}
\Gamma(S\rightarrow  \widetilde X^\dagger \widetilde X ) = \frac{A_{XS}^2}{16\pi m_S} \sqrt{1-\frac{4m_{\widetilde X}^2}{m_S^2}}.
\end{eqnarray}

\begin{figure}[!t]
\centering
{\epsfig{figure=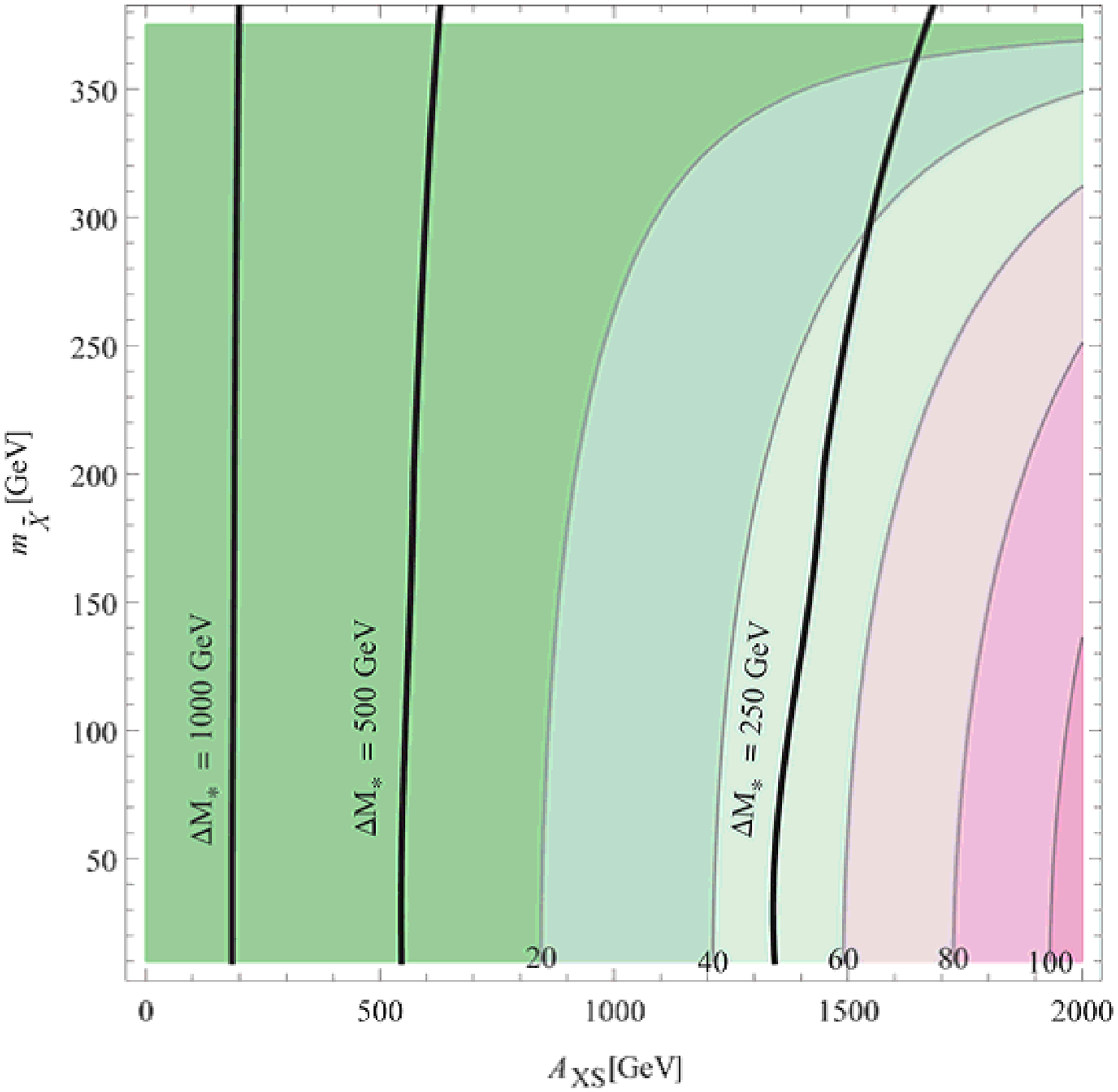,width=0.45\textwidth}}
{\epsfig{figure=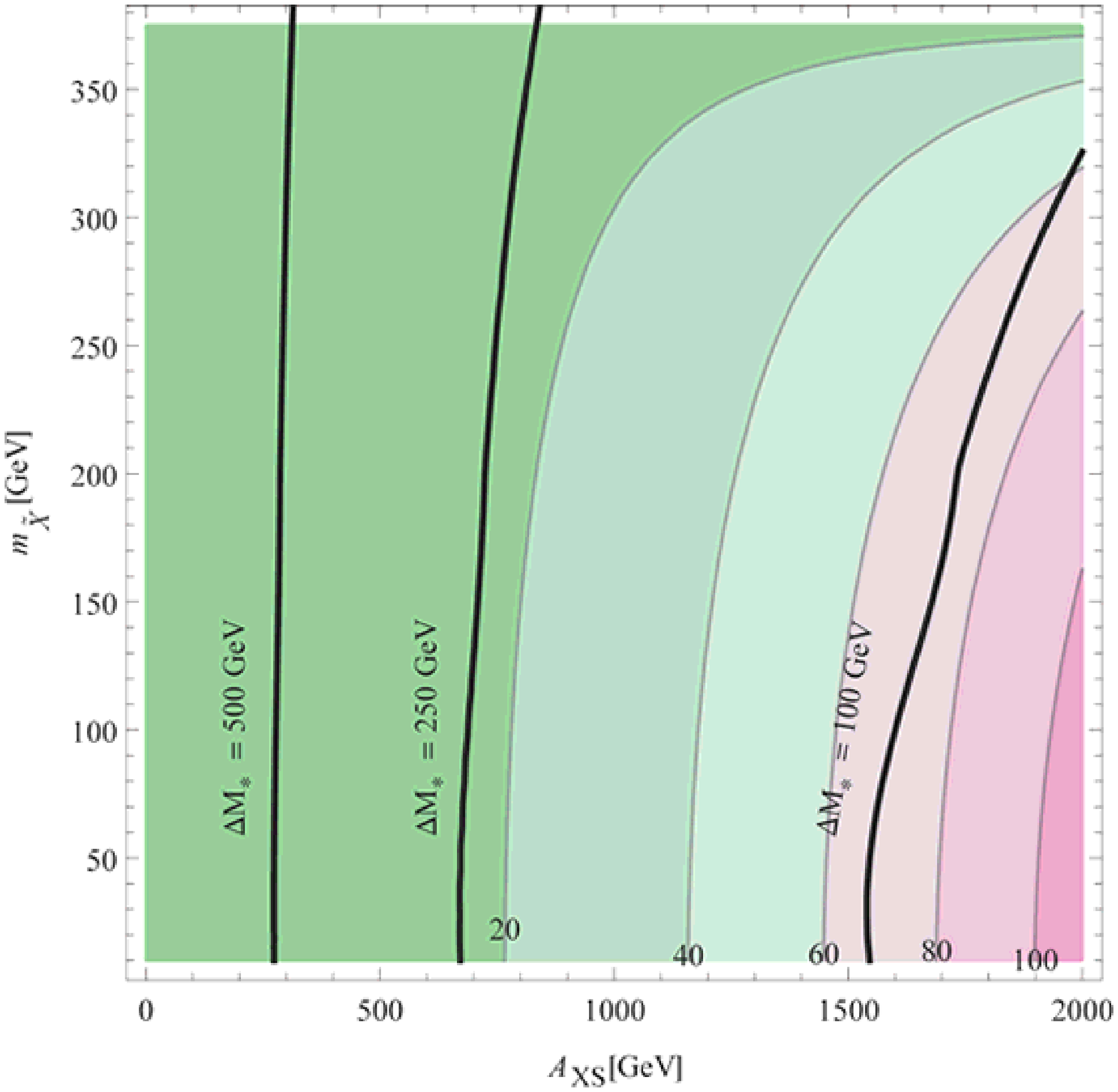,width=0.45\textwidth}}
\caption{
Values of $\Gamma_{\rm tot}$ and comparison with the ATLAS mono-jet search in the type-I model with
$N_u y_U=N_d y_D=3$ (left) and $N_u y_U=N_d y_D=6$ (right). 
Black lines represent the distance against the lower bound on the mass parameter:  
$\Delta M_* \equiv M_* - M_*^{\rm exp}$. 
$M_*$ is defined in Eq.~(\ref{eq:Mstar}).
Background colors depict the total decay width of the scalar resonance 
and attached numbers are its values in the unit of GeV. 
 }
\label{fig2}
\end{figure}

Fig. \ref{fig2} shows $\Delta M_* \equiv M_* - M_{*}^{\rm exp}$ and the total decay width of $S$, 
where $M_{*}^{\rm exp}$ is a lower bound on the effective coupling scale 
can be read from the Fig.10 (f) of Ref.~\cite{Aad:2015zva}. 
Note that the experimental bound on the $M_*^{\rm exp}$ depends on the mass of $ \widetilde X$.  
We see that  $\Delta M_*$ is positive anywhere, 
so the mono-jet search will not give stringent constraints in our models.
The observed best-fit value of the total decay width about $ 45$ GeV can be achieved 
at $A_{XS} \sim 1.5$ TeV without conflict with the mono-jet search.  
\\

Next, we discuss the collider bounds from the new physics searches in $jj$/$b\overline{b}$/$t\overline{t}$ plus
large missing energy channels.
The vector-like quarks should be in the range $300$ GeV $\lesssim m_{U,D} \lesssim 600$ GeV, 
to explain the diphoton excess. 
We could expect that such light exotic particles confront strong bounds from the surveys of the extra colored particles at the LHC. 
In our model, the vector-like quark decays into a SM quark and a $\widetilde{X}$, which is counted as missing energy. 
So the vector-like quarks behave like supersymmetric particles, 
and it induces signals $jj+ E_T^{\rm miss}, bb+E_T^{\rm miss}, tt+E_T^{\rm miss}$.
The decays can be realized by the Yukawa couplings, $(\lambda_{u})_{\alpha i}$ and $(\lambda_{d})_{a i}$,
in Eq. (\ref{eq;VX}). The couplings should be controlled to avoid the strong constraints from flavor physics.
This could be done by imposing a suitable flavor symmetry. 
Our main motivation of this paper is to study the collider bounds, so that
we consider simple alignments of $(\lambda_{u})_{\alpha i}$ and $(\lambda_{d})_{a i}$ 
and then assume the strong bounds from flavor physics are evaded.
Here, we also assume that the new Yukawa couplings are enough small 
that the production process of the vector-like quarks in the t-channel of $ \widetilde X$ exchanging
can be ignored.
If the vector-like quarks are produced by such a process involving non-SM particles, 
exclusion limits will be tightened.

In order to extract exclusion limits for the vector-like quarks, 
we generate UFO model file~\cite{Degrande:2011ua} by using FeynRules~\cite{Alloul:2013bka}.
We use the MadGraph5~\cite{Alwall:2014hca} to simulate signal events with a pair produced vector-like quarks 
at the leading order (LO) with up to an parton.  
The generated events are passed into PYTHIA6~\cite{Sjostrand:2006za} and DELPHES3~\cite{deFavereau:2013fsa} to accomodate parton showering and fast detector simulation. 
The matrix element is matched to parton showers  by the MLM scheme~\cite{Caravaglios:1998yr}. 
The generated hadrons are clustered using the anti-$k_T$ algorithm~\cite{Cacciari:2008gp} with the radius parameter $\Delta R = 0.4$. 
In the analysis for the $bb + E_T^{\rm miss}$ and $tt + E_T^{\rm miss}$ searches, 
we assume b-tagging efficiencies as 77$\%$ and $70\%$, respectively.

\begin{figure}[!tbh]
\centering
{\epsfig{figure=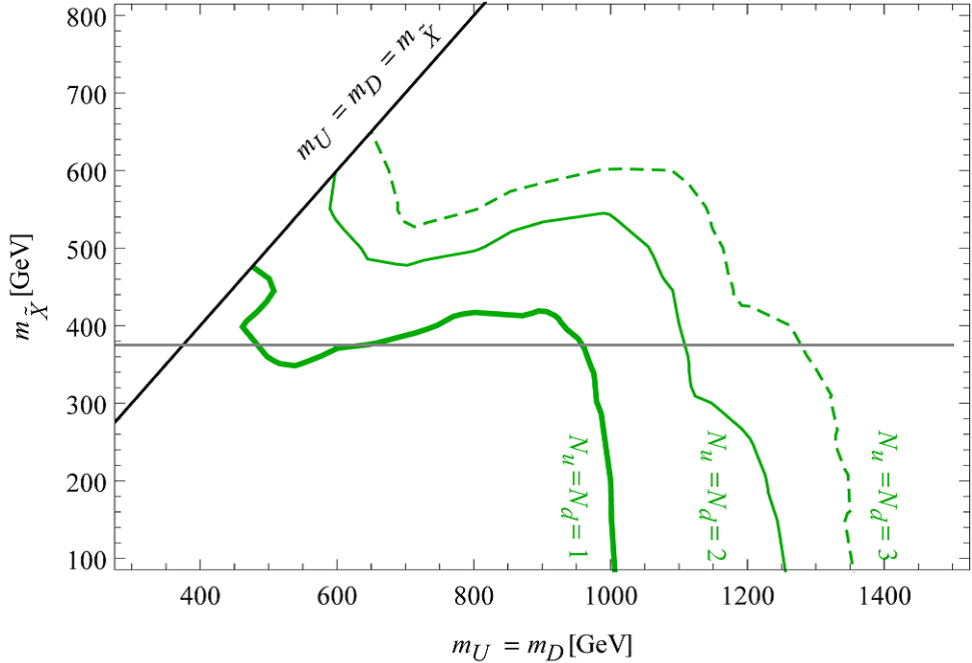,width=0.8\textwidth}}
\caption{
Exclusion limits from $jj + E_T^{\rm miss}$ search. 
The green thick, solid and dashed lines represent the exclusion limits 
for the case with $N_u = N_d = 1,\ 2,\ 3$, respectively. 
The decay of the vector-like quarks to the dark matter $\widetilde{X}$ is kinematically forbidden 
above the black line $m_{U} = m_{\widetilde{X}}$. 
The gray line corresponds to $m_{\widetilde{X}} = 375$ GeV 
which is the threshold whether the scalar resonance can decay into $\widetilde{X}$.
 }
\label{fig3}
\end{figure}

\begin{figure}[!tbh]
\centering
{\epsfig{figure=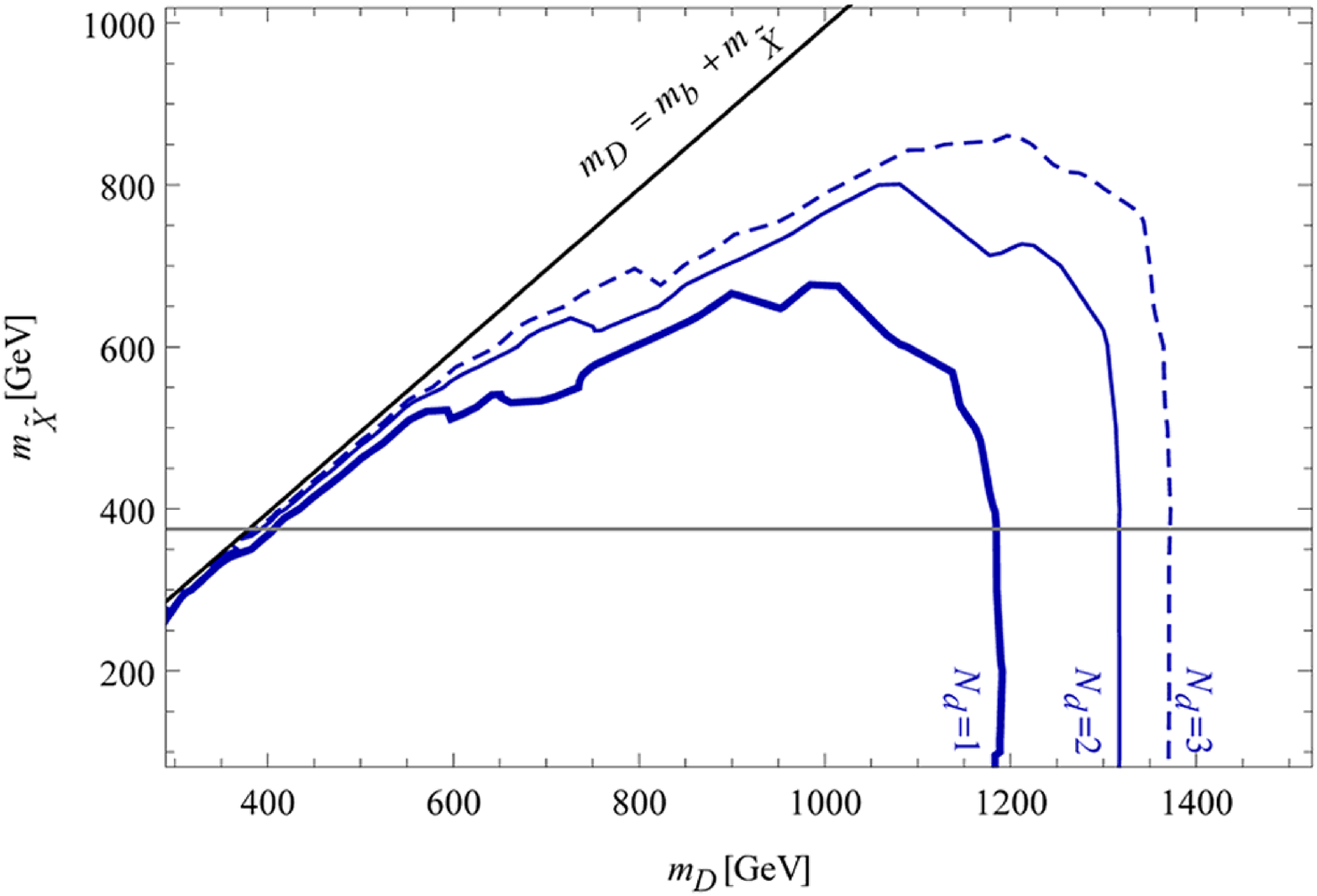,width=0.8\textwidth}}
\caption{
Exclusion limits from $bb + E_T^{\rm miss}$ search. 
The blue thick, solid and dashed lines represent the exclusion limits 
for the case with $N_d = 1,\ 2,\ 3$, respectively. 
The decay of the vector-like quarks to the dark matter $\widetilde{X}$ is kinematically forbidden 
above the black line $m_{D} = m_b + m_{\widetilde{X}}$. 
The gray line corresponds to $m_{\widetilde{X}} = 375$ GeV 
which is the threshold whether the scalar resonance can decay into $\widetilde{X}$.
 }
\label{fig4}
\end{figure}

\begin{figure}[!th]
\centering
{\epsfig{figure=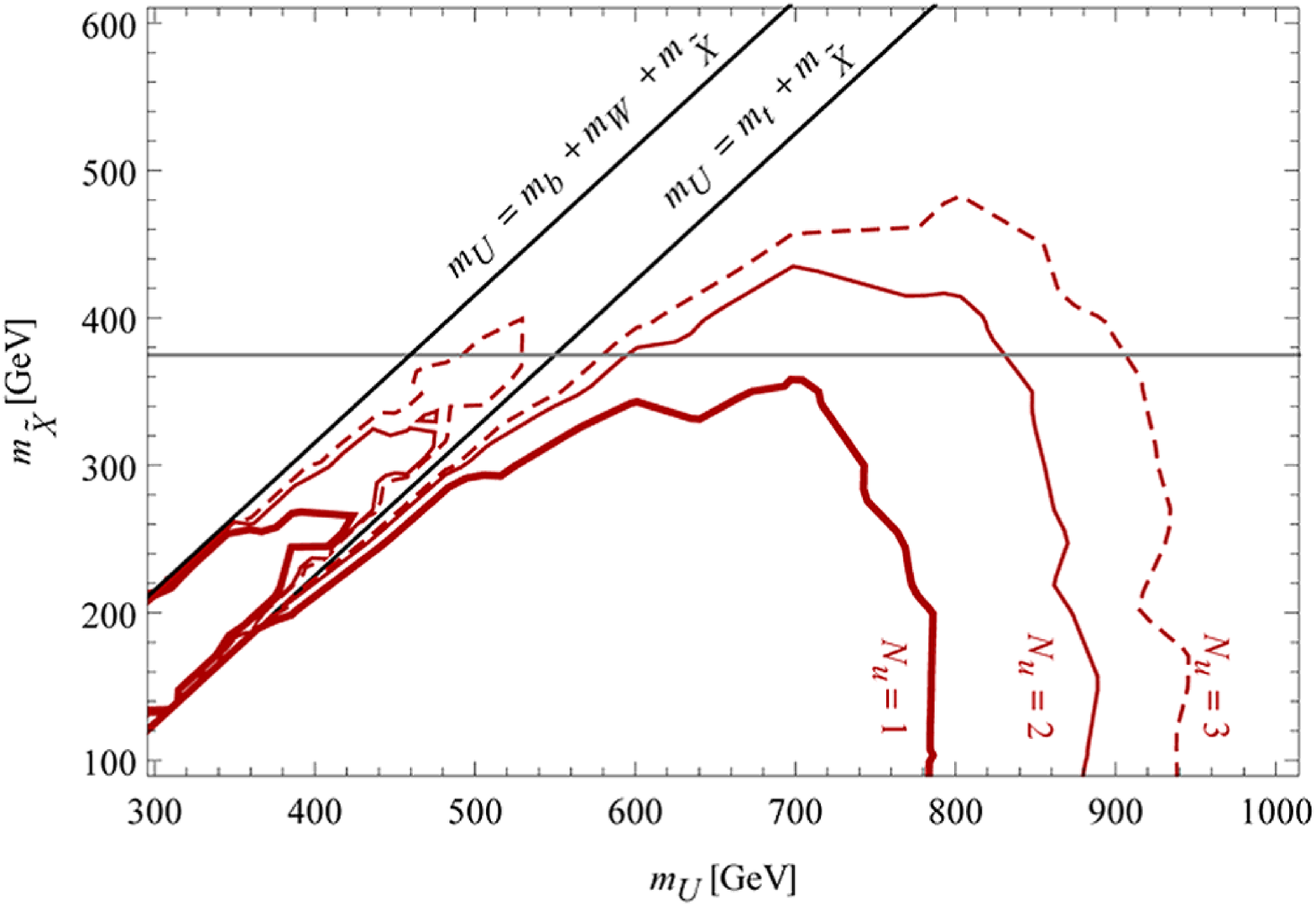,width=0.8\textwidth}}
\caption{
Exclusion limits from $tt + E_T^{\rm miss}$ search. 
The red thick, solid and dashed lines represent the exclusion limits 
for the case with $N_u = 1,\ 2,\ 3$, respectively. 
The two-body (three) decay of the vector-like quarks to the dark matter $\widetilde{X}$ and the SM particles, 
$U \rightarrow t \widetilde{X}$ $(U \rightarrow b W^+ \widetilde{X}$) is kinematically forbidden 
above the upper (lower) black line. 
The gray line corresponds to $m_{\widetilde{X}} = 375$ GeV 
which is the threshold whether the scalar resonance can decay into $\widetilde{X}$.
 }
\label{fig5}
\end{figure}

Figs.\ref{fig3} and \ref{fig4} show exclusion limits 
from the $jj + E_T^{\rm miss}$~\cite{jjMET} and $bb + E_T^{\rm miss}$~\cite{bbMET} searches, respectively. 
These results refer the latest data of the LHC Run-II with $\sqrt{s} = 13$ TeV.
We refer the signal regions named 2jl, 2jm, 2jt for $jj+E_T^{\rm miss}$ search~\cite{jjMET}, 
and all signal regions in the analysis search for $bb + E_T^{\rm miss}$~\cite{bbMET}.   
In the $jj + E_T^{\rm miss}$ channel, $j$ corresponds to up/down quark.
In Fig.\ref{fig3} (Fig. \ref{fig4}), the exclusion limits in the cases with $N_u=N_d=1,\,2,\,3$ $(N_d=1,\,2,\,3)$ are shown as thick, solid, dashed lines, respectively. The extra quarks have degenerate masses in the each case. 
As we see in Figs. \ref{fig3} and \ref{fig4}, most of the parameter space satisfying $300$ GeV $\lesssim m_{U,D} \lesssim 600$ GeV and $m_{\widetilde X} < 375$ GeV has been already excluded by the LHC experiments, 
although the quite narrow region is still alive in Fig. \ref{fig4}
where the mass difference between the down-type vector-like quark and $\tilde{X}$ 
is just above the bottom quark mass. 
Especially, $N_u,\, N_d>1$ is a reasonable choice to evade Landau poles of $y_U$ and $y_D$ appearing at
too low scales. 
However, the lower bounds on $m_U, m_D$ are bigger than 1 TeV.

Fig. \ref{fig5} shows exclusion limits from $tt + E_T^{\rm miss}$ search~\cite{Aad:2014kra} 
obtained in the experiments at the LHC Run-I. 
We refer the signal regions, tN-diag, tN-med, tN-high, bCa-low, bCa-med, 3body in Ref.~\cite{Aad:2014kra},  
which will be sensitive to our vector-like quarks.

In this analysis, we investigate the region where the two or three-body decay is kinematically allowed, 
then $m_{U} > m_b + m_W + m_{\widetilde X}$ is satisfied. 
In the region with $m_{U} < m_{\widetilde X} + m_W + m_b$, 
only the four body decay $U \rightarrow b f f' \widetilde X $ is allowed 
if the vector-like quark only couples to the top quark but not to first- or second-generation up-type quarks. 
In general, it will be difficult to control the flavor structure of the Yukawa couplings, $(\lambda_u)_{\alpha i}$, so that we expect that the two body decays to light quarks and $\widetilde X$ dominate the four-body decay.  
Therefore, the up-type vector-like quark $U$ decays as $U \rightarrow u \widetilde X\ {\rm or}\ c  \widetilde  X$, in the region with $m_{\widetilde X}<m_{U} < m_{\widetilde X} + m_W + m_b$. Then the exclusion limits would be similar to Fig.~\ref{fig3}.  

Finally, we can see the allowed region for the new physics search in the $tt + E_T^{\rm miss}$ channel in Fig. \ref{fig5}. The exclusion limits in the cases with $(N_u=1,\,2,\,3)$ are shown as thick, solid, dashed lines, respectively, assuming the extra quarks are degenerate and dominantly decay to top quarks and $\widetilde X$.  
Again, only the mass degenerate region is still allowed, 
although the bound is weaker than the bounds in Figs. \ref{fig3} and \ref{fig4}. 
Even if we consider the case $m_{ \widetilde X} \le m_S/2$, 
the parameter region where the mass difference $\Delta m \equiv m_{U}-m_{ \widetilde X}$ 
is in $85$ GeV $ \lesssim \Delta m \lesssim 230$ GeV, the DM mass should satify $m_{\widetilde X} \gtrsim 300$ GeV when $N_u =1$. 
While for $N_u = 3$, the mass difference must be in the narrower range: $85$ GeV $\lesssim \Delta m \lesssim 100$ GeV for $m_{\widetilde{X}} \gtrsim 350$ GeV and 
$175$ GeV $\lesssim \Delta m \lesssim 190$ GeV for $m_{\widetilde{X}} \gtrsim 350$ GeV. 
The former is near the threshold of the three-body decay, 
and the latter is near the threshold of the two-body decay. 
Incidentally, there is wider allowed region in $m_{U} \lesssim 600$ GeV and $m_{\widetilde{X}} \ge m_S / 2$, 
although the explanation for the large width by the invisible decay is impossible. 
\begin{table}[!th]
\caption{Input and Output quantities at a sample point. All masses and widths are in GeV and cross section is in fb.
Values of the Yukawa couplings are fixed at $y_U = y_D = 1.0$.}
 \label{table6}
\begin{center}
\begin{tabular}{c|c|c|c|c|c|c|c}
\hline
\hline
input &                   & output & & $N_{\rm signal}$ & & $N_{\rm signal}$& \\ \hline\hline
$N_u=N_d$ &3& $\Gamma (S \rightarrow gg)$          & 1.33 & SRA250 & 5.86 & bCa-med & 6.43 \\
$m_U$& 450& $\Gamma (S \rightarrow \gamma\gamma)$ & 0.00124 & SRA350 & 1.47 & 3body-1 & 0 \\
$m_D$                & 355     & $\Gamma (S \rightarrow XX)$ & 9.52 & SRA450 & 0 & 3body-2 & 4.82 \\
$m_{\widetilde X}$                & 350     &                $\Gamma_{\rm tot}$ & 10.9 & SRB      & 1.47 &  3body-3 & 0 \\
$A_{XS}$&1000 & $\sigma (pp \rightarrow \gamma\gamma)$& 1.00 & bCa-low & 1.61 & 3body-4 & 1.61 \\ \hline \hline
\end{tabular}
\end{center}
\end{table} 
The sample points that can realize 
$\sigma (pp \rightarrow \gamma \gamma) \sim 1$ fb, 
$\Gamma_{\rm tot} \sim 10$ GeV without conflict with the searches for new exotic particles, are shown
in Table.~\ref{table6}.
We know that $jj+E_T^{\rm miss}$ search gives stringent bound even at the mass degenerate region, so that
we consider the case that vector-like quarks couple to only the third generation quarks. 
The first row shows the values of input parameters: the mass parameters and the trilinear coupling $A_{SX}$. 
The second row shows partial and total widths of the singlet $S$, 
and the cross section of the diphoton process. 
The third and fourth rows show the number of events in the relevant signal regions.
The names of signal regions are defined in Refs.~\cite{bbMET,Aad:2014kra} 
and the names of 3body-$1\sim4$ corresponds to each bin of the signal region 3body, 
where $80 (90)$ GeV $< am_{T2} < 90 (100)$ GeV for 3body-1,2 (3,4) 
and $90$ GeV $< m_{T} < 120$ GeV for 3body1,3, $m_T > 120$ GeV for 3body-2,4.
All values are less than the experimental upper bounds.

\subsection{Type-II model}
\label{subsec;type2}
For the type-II model, the new vector-like scalars decay into the invisible fermion $X$ 
and the SM quarks in a same way as squarks. 
Then exclusion bounds on the vector-like scalars would be same as the one for the SUSY particles~\cite{jjMET,bbMET,Aad:2014kra}. 
However, we know that the large number of the vector-like scalar is required to enhance the diphoton signal, 
then the allowed parameter space would be quite narrow.

\section{Conclusion}
\label{sec;conclusion}

In this paper,  
we studied a possibility that the diphoton resonance around 750 GeV, reported by both of the ATLAS and CMS collaborations, is explained by a singlet scalar field produced by the gluon fusion via the loop diagram involving vector-like (scalar) quarks. 
The excess requires 300-600 GeV masses of the extra colored particles to enhance the production and diphoton decay, so we have especially investigated the experimental bounds on such light vector-like (scalar) quarks. 
One interesting feature of this resonance is a relatively large width about 45 GeV. 
This result is not still conclusive, but this may be also an evidence to reveal the feature of the new physics behind the Standard Model. If the wide decay width is settled,
the 750 GeV resonance should decay to new invisible particles 
since the resonance searches by dijet, dilepton or diphoton signals give the stringent bounds. 
Then we consider the extension of the SM model 
introducing a 750 GeV singlet scalar $S$, 
vector-like quarks $U_\alpha$, $D_a$  
and a singlet particle $\widetilde{X}$, which is a dark mater.

In order to explain the diphoton resonance, 
masses of the vector-like (scalar) quarks should be in a range $300$-$600$ GeV, 
where the lower bound comes from the diphoton and the diboson searches at $\sqrt{s}=8$ TeV. 
Besides, we are sure that the vector-like quarks are strongly constrained by the LHC experiments, 
and this is a main subject of this paper. 
The new invisible particle should also be lighter than 375 GeV ($=m_S/2$) 
to make the singlet scalar decay to the dark matters. 
Then we also studied the consistency with the searches for invisible particles.
We conclude that the bounds from the latest ATLAS result are not so severe 
that the ${\cal O}(10)$ GeV decay width cannot be achieved.

In the Type-I model, the signal cross section of the diphoton resonance depends on 
the quantities $y_{U} N_{u}$ and  $y_{D} N_{d}$, 
where $y_U$ and $y_D$ are the Yukawa couplings for the singlet resonance and the vector-like quarks 
and $N_{u,d}$ are the numbers of vector-like quarks. 
In the Type-II model, where the vector-like are squarks, the required values of $y_{\widetilde{U}} \widetilde N_{u}$ and $y_{\widetilde{D}} \widetilde N_{d}$ are ${\cal O}(10)$. 
 This means that $O(10)$ squarks are required as long as we respect the perturbativity of the
 Yukawa couplings. However, such a lot of light squarks can be easily excluded by the latest LHC results
 concerned with the search for supersymmetric particles.

In the Type-I model, the vector-like quarks decay to the dark matter $ \widetilde X$ 
and SM quarks. 
Then the signal pair-produced vetcor-like quarks become $jj\ {\rm or}\ bb\ {\rm or}\ tt+E_T^{\rm miss}$, 
where we assume that vector-like quarks couples to one-type SM quark 
in order to avoid flavor violation. 
Fig.~\ref{fig3}-\ref{fig5} show experimental bounds on the $m_{U}$-$m_{\widetilde{X}}$ planes. 
We can see that most of parameter region satisfying $300$ GeV $\lesssim m_{U} \lesssim$ $600$ GeV and $m_{\widetilde{X}} \le 375$ GeV,
has been excluded. 
However there still remain allowed region where the vector-like quarks and the DM are nearly degenerate, 
when the vector-like quarks decay into one top quark and one DM.

If we consider the case $N_{u} = 1$ that the vector-like quark decays to a top quark with missing energy, 
the region $\Delta m \equiv m_{U} - m_{\widetilde{X}} \lesssim 230$ GeV 
and $m_{\widetilde{X}} \gtrsim 300$ GeV is still allowed. 
Note that as the mass difference decreases, 
the two-body (three-body) decay is kinematically forbidden, 
and then the decay width of the vector-like quark becomes extremely small 
such as $\sim {\cal O}(10^{-(3 {\rm \mathchar`-} 6)}),\ ({\cal O}(10^{-(7 {\rm \mathchar`-}13)}))$ GeV.
In this case, it is generally possible that the two-body decay to light quarks would dominate the decay modes.   
The upper bounds on the mass difference becomes severer as $N_{u}$ increases. 
For instance, the mass difference must be in the range $85$ GeV $\lesssim \Delta m \lesssim 100$ GeV 
or $175$ GeV $\lesssim \Delta m \lesssim 190$ GeV for $N_{u}=3$ 
only when $m_{\tilde{X}} \gtrsim 300$ GeV. 
For the down-type vector-like quarks, the bounds are highly constrained, 
it must be in the range $5$ GeV $\lesssim \Delta m \lesssim 20$ GeV even for $m_{\tilde{X}} \gtrsim 350$ GeV. 
Therefore, the signal cross section of the diphoton signal can reach $O(1)$ fb 
without conflict with the vector-like quarks search, 
when the mass difference is suitably small with the accuracy of at most 15 GeV. 
The favored parameter region can be probed by the upcoming data of the LHC Run-II.

 
\section*{Acknowledgments}
The work of Y.O. is supported by Grant-in-Aid for Scientific research
from the Ministry of Education, Science, Sports, and Culture (MEXT),
Japan, No. 23104011. 


\appendix



\end{document}